\documentclass[11pt]{article}

\usepackage{authblk}

\usepackage{hyperref}

\begin{document}
%






\newcommand{\be}{\begin{equation}}
\newcommand{\ee}{\end{equation}}
\newcommand{\bea}{\begin{eqnarray}}
\newcommand{\eea}{\end{eqnarray}}
\newcommand{\twospinor}[2]{  \left( \begin{array}{ccc}{#1} \\ [6pt] {#2} \end{array} \right) }
\newcommand{\twomatrix}[4]{  \left( \begin{array}{ccc}{#1} & {#2} \\ [6pt] {#3} & {#4} \end{array} \right)  }










\bibliographystyle{elsarticle-num}



%

\title{Unified Picture of Electromagnetic and Gravitational Forces in Two-Spinor Language}


%
\author{Jes\'us Buitrago}


\bigskip

\affil[1]{Faculty of Physics of the University of La Laguna,
38205, La Laguna, Tenerife, Spain}
\affil[2]{jgb@iac.es}

%
%



%


%
%
%

\maketitle
%

%


\begin{abstract}

The well known Geodesic Equation of General Relativity is newly formulated in Weyl two-spinor language in a convenient way susceptible of being combined with a set of two-spinor equations, equivalent to the Lorentz Force of Electrodynamics, obtained in previous studies. General spinor equations of motion for a test charged spin 1/2 particle in arbitrary electromagnetic and gravitational fields fields are obtained describing not only spacetime trajectories but also spin precession. For the case of the Schwarzschild metric and radial trajectories, spinor equations unifying gravitational and electric forces are obtained and solved for a weak gravitational field (or long distance from the source).


\end{abstract} 

%

%

%

\section{Introduction}

In our collective consciousness, we are all aware of the essential role played by the mathematical framework in the formulation of physical laws. In its original form, Maxwell Equations were not easy to understand and even less of unveiling all its content. The adoption of three dimensional vector formalism revealed the close interrelation between electric and magnetic forces. Later, with the advent of Special and General Relativity (GR), the tensor language lead to a unifying picture in terms of the electromagnetic tensor field $F^{\alpha \beta}$ and the underlying Lorentz Invariance of physical laws in flat spacetime as well as general covariance in GR. 

One alternative to tensor calculus is the so called Weyl two-spinor formalism \cite{penrose}. Simply stated, Two-Spinors are elements of a two-dimensional complex vector space. As four-vectors are acted upon by elements of the homogeneous Lorentz Group, two-spinors also change their components when they are acted by elements of the $SL(2,C)$ group (Covering group of the homogeneous Lorentz group). Any tensor equation can be transformed into a spinor equation by certain rules that will be explained in Section 2. However, regarding spinor equations, the opposite statement need not be necessarily true. Given that there are so many tensor equations, to begin with, our attention will address the famous Dirac Equation (although originally not in two-spinor form). The Weyl two-spinor version of the Dirac Equation was obtained in a previous study \cite{bui} fully coincident with Penrose's result in \cite{penrose}:
\begin{equation}
\begin{array}{c}
\nabla^{AA'}\pi _{A}=\frac {m}{\sqrt {2}}\overline\eta^{A'}\\ 
\nabla _{AA'}\overline {\eta }^{A'}=-\frac {m}{\sqrt {2}}\pi _{A}.\\
\end{array}
\end{equation}

Since the rules of spinor algebra are not very familiar among physicists it seems, at least in principle, questionable whether there are any serious reasons for adopting the aforementioned formalism. Notwithstanding, with the devenir of time the concise and elegant Weyl two-spinor language gained adepts  and interest among some physicists and mathematicians specially Roger Penrose who saw in two-spinors a seed for a new insight in physics. Indeed, we share Penrose view "that we have still not yet seen the full significance of spinors (particularly the 2-components ones) in the basis structure of physical laws" \cite{Penrose 1983b}.

%
The approach in this paper to the old issue of the unification of gravitational and electromagnetic interactions is substantially different from what (to my knowledge) have been attempted until now. To begin with, the fundamental variables will not be four-vectors, like momentum or 4-velocity. Instead, in the new picture, they are complex two valued spinors obeying classical spinor equations in which spin 1/2 appears in a natural way via a two dimensional representation of the $SL(2, C)$ group. The replacement of the classical variables in tensor form by two-spinors is a crucial step since it provides a unified language for both classical dynamics and relativistic quantum mechanics. 



%
Perhaps, the reader will note that no mention, or explicit appearance, is made of such important quantities as the Riemann Spinor, Weyl Conformal spinor, and spinor version of Einstein Equations which are already standard material in the literature (\cite{Penrose annals}) and even in textbooks \cite{carmeli}. The Spinor Equations presented in next section and obtained some time ago (\cite{buitrago}) are just the additional  element that makes the Weyl two-spinor language an  alternative (in some cases more convenient) to the conventional tensor calculus.
%

%
%

\section{Spinor Equations and Lorentz force} 

As already mentioned, any tensor equation can be translated to spinor form. In particular the Lorentz Force of Electrodynamics has a well known spinor form \cite{penrose}:
\begin{equation}
	\frac{du_{AA'}}{d\tau}=\frac{e}{m}F_{AA'BB'}u^{BB'}.
\end{equation}
The four-rank antisymmetric spinor is given by

\begin{equation} \label{emfield}
F_{ABA'B'}=\epsilon_{AB}\bar\phi_{A'B'}+
\epsilon_{A'B'}\phi_{AB}.
\end{equation} \\

where $\epsilon_{AB} = \epsilon^{AB}=\epsilon^{A'B'}$ is the spinor metric
\begin{equation}\label{metric1}
\epsilon_{AB}=\left(\begin{array}{cc} 0 & 1 \\ -1& 0 \end{array}\right),
\end{equation} \\
playing a similar role as the metric tensor for raisin (lowering) indices:
$$
\eta^A= \epsilon^{AB}\eta_B, \ \eta_A=\epsilon_{CA}\eta^C.
$$.

The so called symmetric spinor field $\phi_{AB}$ is
\begin{equation} \label{phi}
\phi_{AB}= \frac{1}{2}
\left( \begin{array}{cc}
\left[ E_{1}-iB_{2}\right] -i\left[E_{2}-iB_{2}
\right] & -E_{3}+iB_{3} \\
-E_{3}+iB_{3} & \left [-E_{1}+iB_{2}\right]-i
\left[E_{2}-iB_{2}\right] \\
\end{array}\right) .
\end{equation}
(For the complex conjugate $\bar\phi_{A'B'}$, replace all quantities by their complex conjugates).

The spinor form of the Lorentz Force contains no more information than its tensor counterpart. In order to go a step further in physical insight one needs the following two U(1) local gauge invariant linear first order differential coupled spinor equations which, as in preceding works (see below), are the starting point of this study:
\begin{equation}\label{eqofmot}
\begin{array}{c}
\frac{d\eta^A}{d\tau}=\frac{e}{m}\phi^{AB}\eta_{B} \\
\frac{d\pi_A}{d\tau}=-\frac{e}{m}\phi_{AB}\pi^{B}. \\
\end{array}
\end{equation} \\
(For a short derivation of these equations, in its physical contravariant components, emphasizing its geometrical origin, see \cite{bui3}). As explained in detail in two former studies \cite{buitrago}\cite{bui}, these coupled equations (natural units $\hbar=c=1$ will be used) are equivalent to the Lorentz Force describing the motion of a 1/2 spin particle  of mass $m$ and charge $e$ (typically an electron) under an electromagnetic field described by the symmetric second-rank spinor $\phi_{AB}$,  already written above.
%

%
%
%

%
%

%
%
%

The solution of equations (\ref{eqofmot}) determine the 
four momentum of the particle (or, in this study, the four velocity) given by the hermitian spinor defined  as superposition of the two null directions $\pi^A\bar\pi^{A'}$ and $\eta^A\bar\eta^{A'}$  \cite{bette}:
\begin{equation} \label{momentum}
u^{AA'}=\frac{1}{\sqrt{2}}\left[\pi^{A}\bar\pi^{A'}+\eta^{A}\bar\eta^{A'}\right].
\end{equation} \\
Since $u^{AA'}$ is to represent the four-velocity of a massive particle, must be time-like and certainly fulfill the condition:
\begin{equation} \label{condition}
u^{AA'}u_{AA'}=1,
\end{equation} \\
together with the coupling condition between both spinors: $\eta^A \pi_A=1$ (or the rest mass of the particle $m$ when dealing with the four-momentum).
On the other hand, following the standard representation, the different components of $u^{AA'}$ are labeled according to
\begin{equation} \label{strep}
u^{AA'}=\left( \begin{array}{cc}
u^{00'} & u^{01'}\\
u^{10'} & u^{11'} \\
\end{array}\right)
= \frac{1}{\sqrt 2}\left( \begin{array}{cc}
u^0+u^3 & u^1+i u^2 \\
u^1-i u^2 & u^0-u^3 \\
\end{array}\right).
\end{equation}
As shown in some specific cases in \cite{bui} and \cite{bui3}, this last expression must be used when solving an specific problem to identify the components of $u^{AA'}$ in the solution.


%


\section{Extension to the Gravitational Field}
Previous studies aiming to a unified picture of electromagnetic and gravitational interactions usually depart from the symmetric affine connection via the Cartan Torsion (Any account of the plethora of articles dealing with this subject, beginning with those of Einstein himself, is beyond the scope of this work). However when trying to incorporate the other essential ingredient of normal fermionic mather, namely spin, it is usually necessary to incorporate the Dirac Equation. We have then the origin (during almost the last hundred years) of the so called Einstein-Cartan-Dirac Theories (For a review see \cite{hehl} and a recent study \cite{khanapur}). From a conceptual standpoint, the principal advantage of the approach that will be outlined in this work is that, within the Weyl formalism, spin does appear as something quite natural, already at the classical level, through the two master equations (\ref{eqofmot}) which are the starting point of this and other recent previous studies \cite{bui} \cite{bui3}. Since in the equations of motion, spin is already included, it is only necessary the implementation of a covariant derivative in the left hand side of any of the mentioned equations. To clarify and justify this procedure let us write the conventional tensor geodesic equation modified by the presence of an electromagnetic field:
\begin{equation}\label{geodesic}
	\frac{du^\alpha}{d\tau}+\Gamma^\alpha _{\beta \delta}u^\beta u^\delta = \frac{e}{m}F^\alpha _{\ \beta} u^\beta .
\end{equation}

 From the general covariance principle, the last equation introduces electric charge but spin, the other fundamental property of fermions, is absent. In this way one can say that, in tensor language, the last equation describe the behavior of a test particle without any internal structure other than charge. Many of the studies, during almost the last hundred years, within the so called Einstein-Cartan-Dirac Unifying Theories have attempted to incorporate spin. As we shall see, in our approach, spin is already embedded in the starting equations.

As any tensor equation can be written in spinor form using the machinery of spinor calculus by means of the so called Infeld Van der Waerden symbols
\footnote{The correspondence between spinors and tensors is achieved by means of the mixed quantities (Infeld van der Waerden symbols) $\sigma^\mu_{AB'}$ ($\mu=0,1,2,3$). In flat spacetime they are the Pauli matrices and the unit matrix divided by $\sqrt 2$. For instance, the spinor equivalent of $T_{\mu\nu}$ is given by
\begin{equation}
	T_{AB'CD'}=\sigma^\mu_{AB'}\sigma^\nu_{CD'}T_{\mu\nu}.
\end{equation}}  \cite{infeld},  it is, in principle, possible a literal translation of the former equation to spinor language (note that the affine connection is not a tensor). However, and in view of the precedent comments and according with the standpoint adopted here, it is more convenient to start with equations (\ref{eqofmot}) and generalize them to curved spacetime. 

The spinor covariant derivative is defined as \cite{carmeli}
\begin{equation}
\nabla _{\mu }\xi _{A}=\frac {\partial \xi _{A}}{\partial x^{\mu }}-\Gamma ^{B}_{A\mu }\xi _{B}
\end{equation}

\begin{equation}
	\nabla _{\mu }\xi ^{A}=\frac {\partial \xi ^{A}}{\partial x^{\mu }}+\Gamma ^{A}_{B \mu }\xi ^{B}.
\end{equation}

The spinor affine connection $\Gamma^A_{B\mu}$ can be written in terms of the ordinary tensorial affine connection, the already mentioned Infeld van der Waerden Symbol matrices $\sigma^\alpha_{BB'}$ and their derivatives as \cite{carmeli}
\begin{equation}
	\Gamma^A_{B\mu}=\frac{1}{2}\sigma^{AB'}_\nu \left(\Gamma^\nu_{\mu\rho}\sigma^\rho_{BB'}+\partial_\mu \sigma^\nu_{BB'}\right),
\end{equation}
with a similar expression for the complex conjugate terms.
From the previous expression, contracting with $dx^\mu/d \tau$, the 2-spinor equation extended to curved spacetime counterpart of (\ref{geodesic}) for $\eta^A$ is

\begin{equation} \label{eta}
\frac{d\eta^A}{d\tau}+\frac{1}{2}\sigma^{AB'}_\nu \left(\Gamma^\nu_{\mu\rho}
\sigma^\rho_{BB'}+\partial_\mu \sigma^\nu_{BB'}\frac{dx^\mu}{d\tau}\right)\eta^B
=\frac{e}{m}\phi^A_{\ B} \eta^B,
\end{equation}
with a similar equation for $\pi^A$ and

\begin{equation}
\phi^A_{\ B}= \frac{1}{2}\left[\left(\begin{array}{cc} E_3& E_1+iE_2 \\ E_1-iE_2& -E_3\end{array}\right)+i\left(\begin{array}{cc} B_3& B_1+iB_2 \\ B_1-iB_2& -B_3\end{array}\right)\right].
\end{equation}
\section{Radial trajectories in the Schwarzschild Metric}
In the presence of the gravitational field induced by a mass $M$ and for $\theta =\phi =0$, the metric reduce to
\begin{equation} \label{metric0}
	d\tau^2 = \left(1-\frac{a}{r}\right)dt^2-\frac{1}{1-\frac{a}{r}}dr^2,
\end{equation}
with $\tau$ equal to proper  time and $a=2GM$.

For radial trajectories the non null Christoffel symbols are
\begin{equation}
	\Gamma^0_{01}=\frac{a}{2r^2(1-\frac{a}{r})},\ \Gamma^1_{00}=\frac{1}{2}(1-\frac{a}{r})\frac{a}{r^2}, \ \Gamma^1_{11}=-\frac{a}{2r^2(1-\frac{a}{r})}.
\end{equation}

The Infeld van der Waerden Symbols in curved spacetime, with metric
 $g^{\mu\nu}$ are defined as (In flat spacetime they can be easily obtained from the Pauli matrices plus the unit matrix)
$$
\sigma^\mu_{AB'}\sigma^{\nu AB'}=g^{\mu\nu}.
$$
For the radial part of the metric they are:
\begin{equation}
\sigma^0_{BB'}= \frac{(1-\frac{a}{r})^{-1/2}}{\sqrt 2} \left(\begin{array}{cc} 1 & 0\\ 0 & 1\end{array}\right) \  , \sigma^1_{BB'}= \frac{(1-\frac{a}{r})^{1/2}}{\sqrt 2}\left(\begin{array}{cc} 1 & 0 \\ 0 & -1\end{array}\right).
\end{equation}

\begin{equation}
\sigma_0^{AB'}= \frac{(1-\frac{a}{r})^{1/2}}{\sqrt 2} \left(\begin{array}{cc} 1 & 0\\ 0 & 1\end{array}\right) \  , \sigma_1^{AB'}= \frac{(1-\frac{a}{r})^{1/2}}{\sqrt 2}\left(\begin{array}{cc} 1 & 0 \\ 0 & -1\end{array}\right).
\end{equation}
Although not a difficult task, the calculation of the non null elements of $\Gamma^A_{B\mu}$ is very time consuming (especially if done by hand). In our case they reduce to:
\begin{equation}
	\Gamma^0_{B\mu}: \Gamma^0_{11}=\Gamma^0_{01}=0,\ \Gamma^0_{00}=\frac{1}{4}\frac{a}{r^2}
\end{equation}

and
\begin{equation}
	\Gamma^1_{B\mu}: \Gamma^1_{00}=\Gamma^1_{01}=0,\ \Gamma^1_{10}=-\frac{1}{4}\frac{a}{r^2}.
\end{equation}
After the previous results and equation (\ref{eta}), for the $\eta^A$ spinor, the following equations hold
\begin{equation}
	\frac{d\eta^0}{d\tau}+\frac{1}{4}\frac{a}{r^2}\eta^0 u^0 = \frac{e}{m}\phi^0_B\eta^B
\end{equation}

\begin{equation}
	\frac{d\eta^1}{d\tau}-\frac{1}{4}\frac{a}{r^2}\eta^1 u^0 = \frac{e}{m}\phi^1_B\eta^B.
\end{equation}

The time component of the four velocity $u^0$ is related to $u_0=\widetilde E$ (a conserved quantity that can be interpreted as energy per unit mass \cite{chandra}) by 
\begin{equation} \label{u0}
	u^0=\frac{\widetilde E}{1-\frac{2GM}{r}},
\end{equation}
where we have made the substitution $a=2GM$ (In natural units $G=6.7071\times 10^{-45} MeV^{-2}$. Distances will be measured in $MeV^{-1}$).

To examine the content of the last two equations for $\eta^A$ it is convenient to write the electromagnetic field spinor $\phi^A_{\ B}$ in spherical coordinates. For radial trajectories and taking the $z$ coordinate in the radial direction the only components that need to be considered are $E_r$ and $B_r$:

\begin{equation}
\phi^A_{\ B}= \frac{1}{2}\left[\left(\begin{array}{cc} E_r & 0 \\ 0 & -E_r\end{array}\right)+i\left(\begin{array}{cc} B_r& 0 \\ 0 & -B_r\end{array}\right)\right].
\end{equation}
%
First consider the case in which we only have a point charge source of mass $M$ and positive charge $Q$ and eventually also a radial magnetic field $B_r$. With the above substitutions, the $\eta^A$ spinor equations can be written as:
\bigskip
\begin{equation} \label{fund0}
\frac {d\eta ^{0}}{d\tau }+\frac {1}{2m}\left[ \frac {GMm}{r^{2}}\frac {\widetilde E}{1-\frac {2GM}{r}}-\frac {eQ}{r^{2}}-ieB_r \right] \eta ^{0}=0
\end{equation}
\begin{equation} \label{fund1}
\frac {d\eta ^{1}}{d\tau }-\frac {1}{2m}\left[ \frac {GMm}{r^{2}}\frac {\widetilde E}{1-\frac {2GM}{r}}-\frac {eQ}{r^{2}}-ieB_r \right] \eta ^{1}=0.
\end{equation}
The first thing to note is that in the term corresponding to the gravitational interaction, according to the equivalence principle, the mass $m$ of the test particle cancel out. For a radial distance $r \gg 2GM$ the attraction force reduces to the Newtonian one increasing and becoming singular at $r=2GM$. In the second term we see that Coulomb inverse square law is valid for all distances and that electromagnetic forces depend on the ratio charge/mass. If instead of the electric field $E_r$ corresponding to a charge $Q$ we have a magnetic field in the radial direction $B_r$, then we find  $\pi^A$ and $\eta^A$ eigenfunctions of the spin operator $S_3$ with eigenvalues $\pm 1/2$ respectively and the typical spin precession with frequency $\omega=eB/m$ corresponding to a g-factor 2 of a spin 1/2 particle (see \cite{bui}).
In general, electric and magnetic forces are overwhelmingly dominating (building the world around us) except for macroscopic bodies and scenarios where the total charge averages to zero.

In what follows, the solutions for $r \gg 2GM, \widetilde E=1$ and without any magnetic field, will be found. From (\ref{u0}) and the metric (\ref{metric0})
\begin{equation} \label{dtaudr}
	\frac{d\tau}{dr}=\frac{1}{\sqrt{\widetilde E^2-(1-\frac{2GM}{r}}}.
\end{equation}

For $\widetilde E=1$, equations (\ref{fund0}) and (\ref{fund1}) reduce to
\begin{equation}
\frac {d\eta ^{0}}{d\tau }+\frac {1}{2m}\left[ \frac {GMm}{r^{2}}-\frac {eQ}{r^{2}}\right] \eta ^{0}=0
\end{equation}
\begin{equation}
\frac {d\eta ^{1}}{d\tau }-\frac {1}{2m}\left[ \frac {GMm}{r^{2}}-\frac {eQ}{r^{2}}\right] \eta ^{1}=0.
\end{equation}
By an easy change of variable, using (\ref{dtaudr}), one obtains
\begin{equation}
	\frac{d\eta ^{0}}{dr}+\frac{1}{2m}\sqrt{\frac{r}{2GM}}\cdot \frac{K}{r^{2}}\eta ^{0}=0
\end{equation}
\begin{equation}
\frac{d\eta ^{1}}{dr}-\frac{1}{2m}\sqrt{\frac{r}{2GM}}\cdot \frac{K}{r^{2}}\eta ^{1}=0,
\end{equation}
with $K=GMm-eQ$.

These equations can be easily solved for $r$, the solutions are
%
\begin{equation}
	\eta ^{0}=C_{1}\cdot \exp \left( \frac{1}{m}\frac{K}{\sqrt{2GM}}\cdot \frac{1}{r^{1/2}}\right) 
\end{equation}
and
\begin{equation}
	\eta ^{1}=C_{2}\cdot \exp \left( -\frac{1}{m}\frac{K}{\sqrt{2GM}}\cdot \frac{1}{r^{1/2}}\right) 
\end{equation}
With a similar expression for $\pi^A$.

The procedure of obtaining the usual dynamical variables ($p^\mu, u^\mu$) is always the same and have been ilustrated in previous works (see Ref.[8]). For $C_1=C_2=1/2$ the solutions are
\begin{equation}
	u^{0}=\cosh \left( 2B\frac{1}{r^{1/2}}\right)
\end{equation}

\begin{equation}
	u^{1}=\sinh \left( 2B\frac{1}{r^{1/2}}\right)
\end{equation}

Satisfying the condition
$$
(u^0)^2 - (u^1)^2 = 1.
$$

Series expansion of both components of the four velocity leads to:
\begin{equation}
	u^0=1+\left( 2B \frac{1}{r^{1/2}}\right) ^2 ....
\end{equation}
\begin{equation}
	u^1= \left( 2B \frac{1}{r^{1/2}}\right) +\left( 2B \frac{1}{r^{1/2}}\right)^3 ...
\end{equation}

With $B$ equal to
\begin{equation} \label{B}
B=\frac{1}{\sqrt{2}}\frac{1}{m}\left[ \sqrt{GM}m-\frac{eQ}{\sqrt{GM}}\right] 
\end{equation}
(Notice the mixing coupling terms involving $G$ and $Q$).

For $Q=0$ the second equation leads to the newtonian expression of the scape velocity
$$
v=\sqrt \frac{2GM}{r}.
$$
For $Q \ne 0 $, in some cases, the series expansion diverges even if the condition $r \gg 2GM$ holds, emphasizing the overwhelming preponderance of electromagnetic forces.

Given that $G=6.70711\times 10^{-45} MeV^{-2}$, the mixing second term is negligible in normal astronomical conditions. For instance, for a test particle of the mass of the Earth in radial fall towards the Sun from a distance equal to an AU, from (37), the second term is
\begin{equation}
	\frac{-2eQ}{m\sqrt{2GM.r}}=\frac{-2eQ}{4.06158\times {10^{48}}},
\end{equation}
and
$$
2GM=15203.7 MeV^{-1} \ , r=7.60185\times 10^{11} MeV^{-1},
$$
so that $r \gg  2GM$.
One should notice that small as the quotient in the last equation  seems it could be significative if both bodies had a certain amount of free charges in their surfaces, or within them, but the Universe would be an explosive one in such case.

\bigskip

\section{Conclusions}
We have seen how the two-spinor formalism via the generalization of equations (\ref{eqofmot}) to curved spacetime provides a unifying picture of both gravitational and electromagnetic interactions including spin. In the special case of radial trajectories, gravitational and electric forces are approached in a new integrating picture in which mass, charge and spin are accounted for. It is shown that while Coulomb law is valid at all distances, Newton gravitational inverse square law remains valid only for weak fields or long distances (i.e. for $r\gg 2GM$), increasing and reaching a singularity when $r=2GM$. When there is a magnetic field in the radial direction, a test charged particle with spin one-half precesses internally with a frequency that previously was only accounted for within the contest of relativistic quantum mechanics.
\bigskip
%

%

%

%

\end{document}